\documentclass[prl,amsmath,amssymb,twocolumn,superscriptaddress]{revtex4-2}
\usepackage{graphicx}
\usepackage{subfigure}
\usepackage{adjustbox}
\usepackage{bm}

\usepackage{bbm}
\usepackage{color}
\usepackage{braket}
\usepackage{standalone}
\usepackage{multirow}
\usepackage{tikz}
\usepackage{mathrsfs}
\usepackage{dsfont}
\usepackage[colorlinks,bookmarks=true,citecolor=blue,linkcolor=blue,urlcolor=blue]{hyperref}
\usepackage{cleveref}
\usepackage{comment}
\usepackage{mathtools}
\usepackage{soul}
\usepackage{orcidlink}

\makeatletter
\newcommand{\tableofsmcontents}{%
  \section*{CONTENTS}
  \@starttoc{smt}
}

\newcommand{\smsection}[1]{%
  \section*{#1}
  \addcontentsline{smt}{section}{#1}
}

\makeatother

\begin{document}

\title{Quantum spin Hall crystals at fractional filling of twisted MoTe$_2$}

\author{Xiaoyang Shen\orcidlink{0009-0009-7708-9254}}
\affiliation{Department of Physics, Tsinghua University, Beijing, 100084, China}
\affiliation{Department of Physics, Stockholm University, AlbaNova University Center, 106 91 Stockholm, Sweden}

\author{Raul Perea-Causin\orcidlink{0000-0002-2229-0147}}
\affiliation{Department of Physics, Stockholm University, AlbaNova University Center, 106 91 Stockholm, Sweden}\affiliation{Nordita, KTH Royal Institute of Technology and Stockholm University, 106 91 Stockholm, Sweden}
\author{Christopher Ekman\orcidlink{0009-0003-6426-2376}}
\affiliation{Department of Physics, Stockholm University, AlbaNova University Center, 106 91 Stockholm, Sweden}
\author{Jiong-Hao Wang\orcidlink{0000-0002-4161-9614}}
\author{Hui Liu\orcidlink{0009-0009-4988-9561}}\thanks{hui.liu@fysik.su.se}
\affiliation{Department of Physics, Stockholm University, AlbaNova University Center, 106 91 Stockholm, Sweden}

\author{Emil J. Bergholtz\orcidlink{0000-0002-9739-2930}}
\affiliation{Department of Physics, Stockholm University, AlbaNova University Center, 106 91 Stockholm, Sweden}

\date{\today}

\begin{abstract}
We predict and classify interaction-driven {\it quantum spin Hall crystals} (QSHCs), a class of states emerging at fractional filling through an interplay of topology and spontaneous translation-symmetry breaking. QSHCs form nearly degenerate manifolds whose members can realize distinct topological phases protected by time-reversal or valley $U(1)_v$ symmetry, with time-reversal acting nontrivially within the manifold.  As a representative of this broad class of states we provide evidence for 9-fold quasi-degenerate $\sqrt{3}\times\sqrt{3}$ charge ordered QSHCs at $\nu = -8/3$ of twisted bilayer MoTe$_2$ near a $5^\circ$ twist. Here a $\mathbb{Z}_3$ index organizes states related by lattice translation into three time-reversal invariant states with nontrivial $\mathbb{Z}_2$ topology and three time-reversal related doublets whose individual members spontaneously break time-reversal and carry a $U(1)_v$ protected spin-Chern number.
Finally, we determine the conditions that favor QSHCs over closely competing intervalley-coherent crystals.

\end{abstract}

\maketitle

\emph{Introduction.}---%
Moiré superlattices provide a versatile platform for realizing correlated and topological quantum matter~\cite{andrei2020graphene,andrei2021marvels,mak2022semiconductor,Li2026QuantumPhases}.
A particularly striking example is twisted bilayer MoTe$_2$, where the fractional quantum anomalous Hall effect~\cite{kol_read,andrei_fci,ParticleHoleDuality_Abouelkomsan_2020,patrick_fci,ChernBandsTwisted_Repellin_2020,FractionalQuantumAnomalous_Reddy_2023,SpontaneousFractionalChern_Li_2021,FractionalChernInsulator_Wang_2024} has been observed without an external magnetic field~\cite{SignaturesFractionalQuantum_Cai_2023,ObservationFractionallyQuantized_Park_2023,ThermodynamicEvidenceFractional_Zeng_2023,ObservationIntegerFractional_Xu_2023}. 
At small twist angles, the same platform hosts a broad family of correlated phases, including integer Chern insulators, correlated insulators, generalized Wigner crystals and doping-induced chiral superconductivity~\cite{fan2025superconductivity,xucheng_chiralsuperconductivity}. 
Much of this phenomenology is associated with spontaneous valley-polarization and the concomitant breaking of time-reversal symmetry (TRS). 
By contrast, the role of TRS-preserving topology has received comparatively less attention, even though TRS gives rise to equally rich physics of its own. In the large twist angle regime of twisted MoTe$_2$, the integer filling $\nu=-2$ of holes hosts a quantum spin Hall insulator (QSHI)~\cite{EvidenceFractionalQuantum_Kang_2024,TopologicalInsulatorsTwisted_Wu_2019,jin2026observationmottquantumspin,InteractionDrivenTopological_Qiu_2023,GateTunableAntiferromagnetic_Liu_2024,MagicTwistedTransition_Devakul_2021,InterplayTopologyCorrelations_Xu_2025}, a genuinely TRS-protected topological phase characterized by a $\mathbb{Z}_2$ invariant~\cite{kanemele,qshi,RevModPhys.88.035005}. 

Correlations can further intertwine topology with spontaneous translation-symmetry breaking: a quantum anomalous Hall crystal combines charge order with a quantized Hall response at non-integer fillings. 
In continuum systems, the charge order spontaneously breaks continuous translation symmetry~\cite{halperin_crystal,AnomalousHallCrystals_Dong_2024,AnomalousHallCrystals_Tomohiro_2024,TheoryQuantumAnomalous_Dong_2024,ParentBerryCurvature_Tan_2024,ahc_boran,QuantumAnomalousSpin_Kudo_2024,NonAbelianChern_Uchida_2026}, whereas in a moiré system it may instead break the discrete moiré translation symmetry by enlarging the superlattice unit cell
~\cite{QuantumAnomalousHall_Sheng_2024,PereaCausin2025,Su2025,Reddy2026,liu2025fractionalcherninsulatorstopological,2026arXiv260316374W,Chen_2026}.
Across these scenarios, the resulting phase is equivalent to a Chern insulator possessing chiral edge states, with the defining feature of coexisting spontaneous crystalline order and quantized anomalous Hall response. 
Together, these two threads pose a natural question: can an analogous charge-ordered state support the quantum spin Hall effect at fractional filling? 

\begin{figure}
    \centering
    \includegraphics[width=0.8\linewidth]{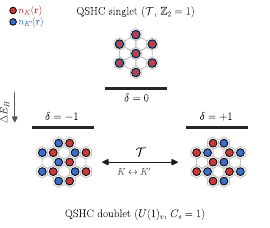}
    \caption{
$\sqrt{3}\times \sqrt{3}$ QSHC structures.
Red and blue circles denote the $K$- and $K'$-valley density centers. The $\delta=0$ sector preserves time reversal ($\mathcal T$), while $\delta=\pm1$ are exchanged by $\mathcal{T}$ and form a degenerate doublet. The levels illustrate the intervalley-Hartree splitting $\Delta E_\text{H}$.
}
    \label{fig:QSHC_trio}
\end{figure}

In this Letter, we answer this question by predicting interaction-driven quantum spin Hall crystals, a class of crystalline phases exhibiting
quantum spin Hall signatures at fractional filling through an interplay
between TRS and discrete translation symmetries. In QSHCs, interaction-induced crystal order gaps the Fermi surface and yields
minibands with opposite topological character in the two spin-valley ($K$ and $K^\prime$) sectors.
TRS further organizes the members of the QSHC family according to their protecting symmetries,
as shown in Fig.~\ref{fig:QSHC_trio}. As a concrete realization, we find $\sqrt{3}\times\sqrt{3}$ QSHCs at fractional filling $\nu = -8/3$ of twisted MoTe$_2$ near a $5^\circ$ twist at the mean-field level. Their ground state manifold contains both TRS-invariant
$\mathbb Z_2$ states and TRS-breaking spin Chern states protected by
valley $U(1)_v$, exemplifying the QSHC family structure described above, cf. Fig.~\ref{fig:QSHC_trio}. In addition,, we find an intervalley-coherent (IVC) crystal competing closely with QSHCs, which gives way to the QSHC upon reducing the effective interaction strength. Finally, we establish a classification of QSHC families, offering a unified framework for the crystal phases with quantum spin Hall character.

\begin{figure*}[t]
    \centering
    \includegraphics[width=1\linewidth]{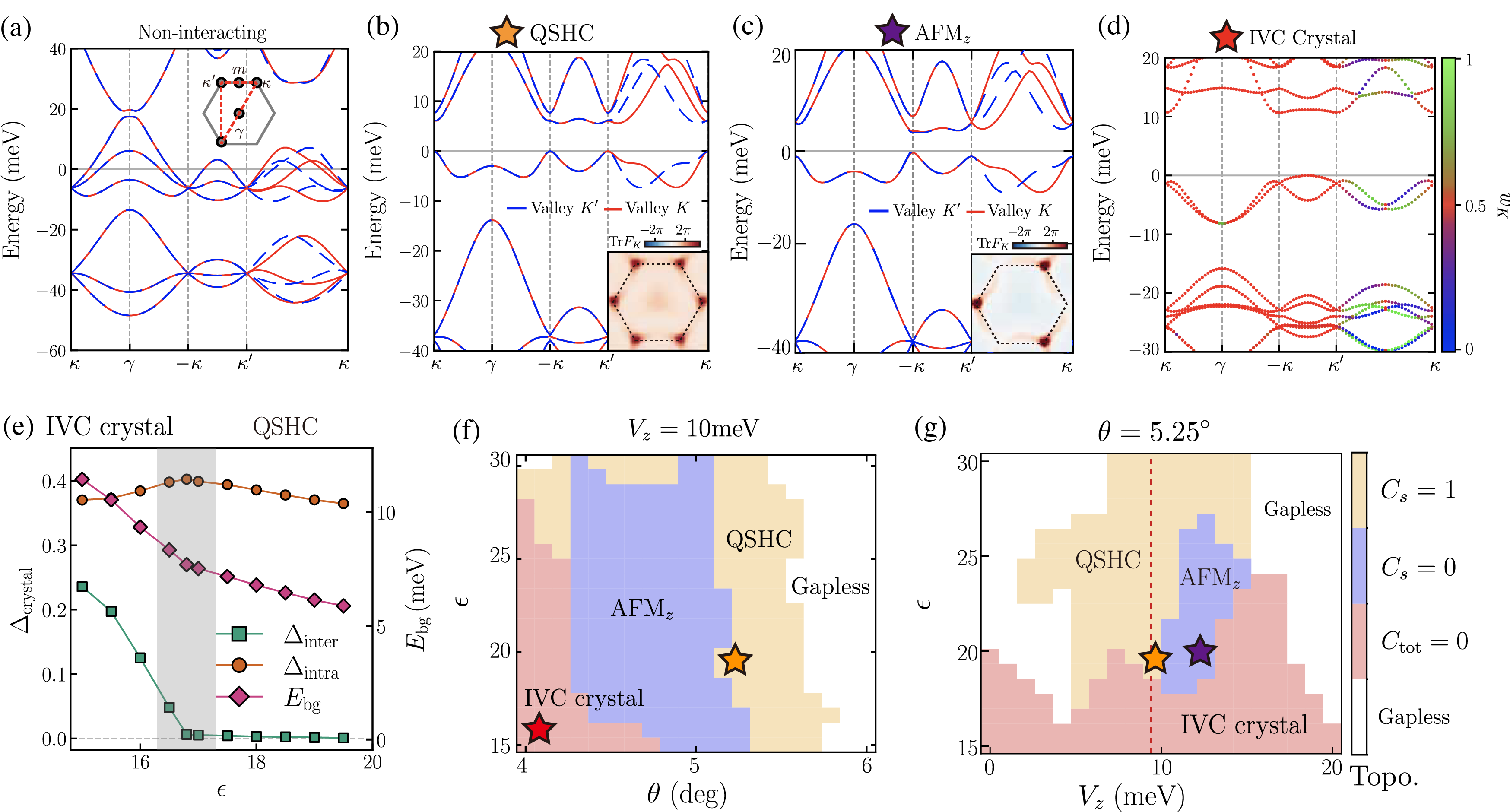}
    \caption{
    (a--d) Crystal band structures of the non-interacting system, the QSHC, the AFM$_z$ state, and the IVC crystal, respectively. For the QSHC and AFM$_z$ states, the valley-resolved non-Abelian Berry curvature $\operatorname{Tr}F_K$ of the occupied bands is shown in the insets. For the IVC crystal, the bands are colored by the valley weight $w_K$. (e) The IVC crystal--QSHC phase transition at $V_z = 9.5$~meV and $\theta = 5.25^\circ$ (red dashed line in (g)), characterized by the indirect HF band gap $E_{\text{bg}}$ and the crystal order parameters strength $\Delta_{\text{crystal}} = \sum_{\boldsymbol{k}}(\operatorname{Tr}\tilde{P}(\boldsymbol{k})^2)^{1/2}/(2N_{\boldsymbol{k}})$, where $\tilde{P}$ denotes the traceless part of the intervalley (intravalley) block of the density matrix for the intervalley (intravalley) crystal channel. (f,g) Phase diagrams in the $\epsilon$--$\theta$ plane at $V_z = 10$~meV and in the $\epsilon$--$V_z$ plane at $\theta = 5.25^\circ$. The QSHC and AFM$_z$ phases are labeled by their spin-Chern numbers, while the IVC crystal carries a vanishing total Chern number. The red dashed line in (g) corresponds to the trajectory in (e).}
    \label{fig:phase_diagram}
\end{figure*}

\emph{Non-interacting spectrum.}---%
We start from the non-interacting continuum model of
twisted MoTe$_2$ \cite{TopologicalInsulatorsTwisted_Wu_2019,FractionalChernInsulator_Wang_2024} at filling $\nu = -8/3$ (see Supplemental Material, SM~\cite{supplemental}).
To investigate the
possible commensurate crystallization, we fold the band structure from the moiré Brillouin zone (mBZ) into the
Brillouin zone of the $\sqrt{3}\times\sqrt{3}$ superlattice, which we call
the crystal Brillouin zone (cBZ), shown in Fig.~\ref{fig:phase_diagram}(a). At
$\nu = -8/3$ the non-interacting system is metallic, with a finite Fermi
surface. 
Moreover, the folded bands exhibit Dirac-cone-like crossings at the cBZ
corners $\kappa$ and $\kappa'$, which are protected by the moiré
translation symmetry. Below we will show that interactions gap out these Dirac cones and break the original moiré translation symmetry, forming a $\sqrt{3}\times\sqrt{3}$ crystal order with rich symmetries and topology.

\emph{QSHC and competing phases.}---%
We now search for topological crystals induced by interactions. 
Typically, distinct crystalline orders can be distinguished via the order parameters:
\begin{equation}
\Phi_\mu(\boldsymbol{Q}) = \frac{1}{N_{}}
\sum_{\boldsymbol{k}\in\mathrm{mBZ}}
\langle c^\dagger_{\boldsymbol{k}+\boldsymbol{Q}}\,\hat{\tau}_\mu\,
c_{\boldsymbol{k}}\rangle ,\quad \mu = 0,x,y,z.
\label{eq:order}
\end{equation}
with $\boldsymbol{Q}$ a $\sqrt{3}\times\sqrt{3}$ reciprocal superlattice
vector for band-folding and $\hat{\tau}_\mu$ the Pauli matrices in valley space, $N$ is the number of momentum points.
$\Phi_0$ and $\Phi_z$ are the intravalley charge and
spin channels, respectively, and $\Phi_{x,y}$ is the intervalley coherent (IVC) channel. We perform momentum-space self-consistent Hartree-Fock (SCHF) calculations in the cBZ,
retaining both $K$ and $K^\prime$ valleys and projecting onto the active crystal
minibands,
with twist angle $\theta$, dielectric constant $\epsilon$, and displacement field $V_z$ as tuning parameters (see the SM~\cite{supplemental} for numerics).

Fig.~\ref{fig:phase_diagram}(f) shows the SCHF phase diagram in the
$\theta$-$\epsilon$ plane at $V_z = 10~\mathrm{meV}$, where we find
four phases in the large twist angle regime ($\theta \in[4^\circ,6^\circ]$). 
The first is the QSHC shown in Fig.~\ref{fig:phase_diagram}(b), dominated by the
intravalley channel $\Phi_0$. 
Its valley Chern numbers $C_K = -C_{K'} = 1$, are obtained from the
 Berry curvature $\operatorname{Tr}\!F_\tau$ of the occupied bands in each valley~\cite{fukuichern}, leading to a spin-Chern number $C_s \equiv (C_K - C_{K'})/2 = 1$. 
The Berry curvature shown in the inset of Fig.~\ref{fig:phase_diagram}(b), concentrating at
$\kappa$ and $\kappa'$, indicates that $\Phi_0$ gaps out the crystal Dirac cones of the non-interacting crystals. The second phase is the AFM$_z$
crystal in Fig.~\ref{fig:phase_diagram}(c), which carries
$C_K = C_{K'} = 0$ and a strong N\'eel-like
out-of-plane spin order $\Phi_{z}$. It differs from QSHC by the absence of quantized helical edge current, as the crystal Dirac cones at $\kappa$ and $\kappa^\prime$ are distinctly gapped and contribute differently to topology, leading to a trivial crystal, as shown in the inset of Fig.~\ref{fig:phase_diagram}(c). The third phase is the topologically trivial intervalley-coherent (IVC) crystal shown in
Fig.~\ref{fig:phase_diagram}(d), characterized by a finite intervalley
order parameter $\Phi_{x,y}$. The dominant IVC order occurs at the
$\sqrt{3}\times\sqrt{3}$ ordering wave vector and produces a
finite-momentum in-plane spin texture reminiscent of the
$\sqrt{3}\times\sqrt{3}$, $120^\circ$ in-plane antiferromagnet at
$\nu=-2$ \cite{InteractionDrivenTopological_Qiu_2023,DiverseMagneticOrders_Wang_2023}. The IVC crystal also develops a
finite $\Phi_0$, which contributes to its intravalley crystal order.
Finally, the gapless regime comprises states with a vanishing Hartree--Fock indirect band gap $E_{\text{bg}}$, with or without additional symmetry breaking.

The QSHC occupies an intermediate window of twist angle,
$\theta \simeq 5\text{--}5.5^\circ$, giving way to the IVC crystal at
smaller $\theta$ and, at larger $\theta$
, to a metal as the interaction-driven gap closes and the indirect band gap $E_{\text{bg}}$ vanishes.
The IVC-QSHC transition reflects the competition between different crystal channels. As shown in Fig.~\ref{fig:phase_diagram}(e), in IVC crystal phases, the intravalley and intervalley channels coexist to support the crystal order. When decreasing the interaction or increasing the bandwidth, the intervalley channel decays, and first vanishes at a critical value $\epsilon_c\approx 16.8$. By contrast, the intravalley component continues
to grow before the transition to QSHC and sustains the indirect band gap
. We emphasize that $\epsilon$ parametrizes the effective screening within our projected SCHF description; screening from additional degrees of freedom and correlations beyond Hartree--Fock can quantitatively renormalize the phase boundaries.
After the intervalley block vanishes, the $U(1)_v$ symmetry (valley-charge conservation) is restored, and the intravalley channels dominate the crystal order, signaling the onset of the QSHC phase. The HF calculations suggest the IVC-QSHC phase transition could be a weakly first-order or continuous phase transition, which we leave for future study.


We note that a limitation of SCHF is its inability to capture fractionalized topological orders. 
Nevertheless, even in the valley polarization limit, the fractional Chern insulators reported earlier transition into a gapless phase in the large-twist-angle regime where the QSHC is stabilized. Moreover, two-valley exact diagonalization shows that the ground state in this regime is not valley polarized (see SM~\cite{supplemental} for details). 
This further suggests that the valley-polarized generalized Wigner crystals and quantum anomalous Hall crystal found at small twist angle and different fillings are not direct competitors of the QSHC~\cite{hmgc-shx3,GateTunableAntiferromagnetic_Liu_2024,InteractionDrivenTopological_Qiu_2023,DiverseMagneticOrders_Wang_2023,ObservationFerromagneticPhase_An_2025}.
These facts justify our focus on time-reversal-related crystal orders as the principal competing states.


\begin{figure}
    \centering
    \includegraphics[width=0.9\linewidth]{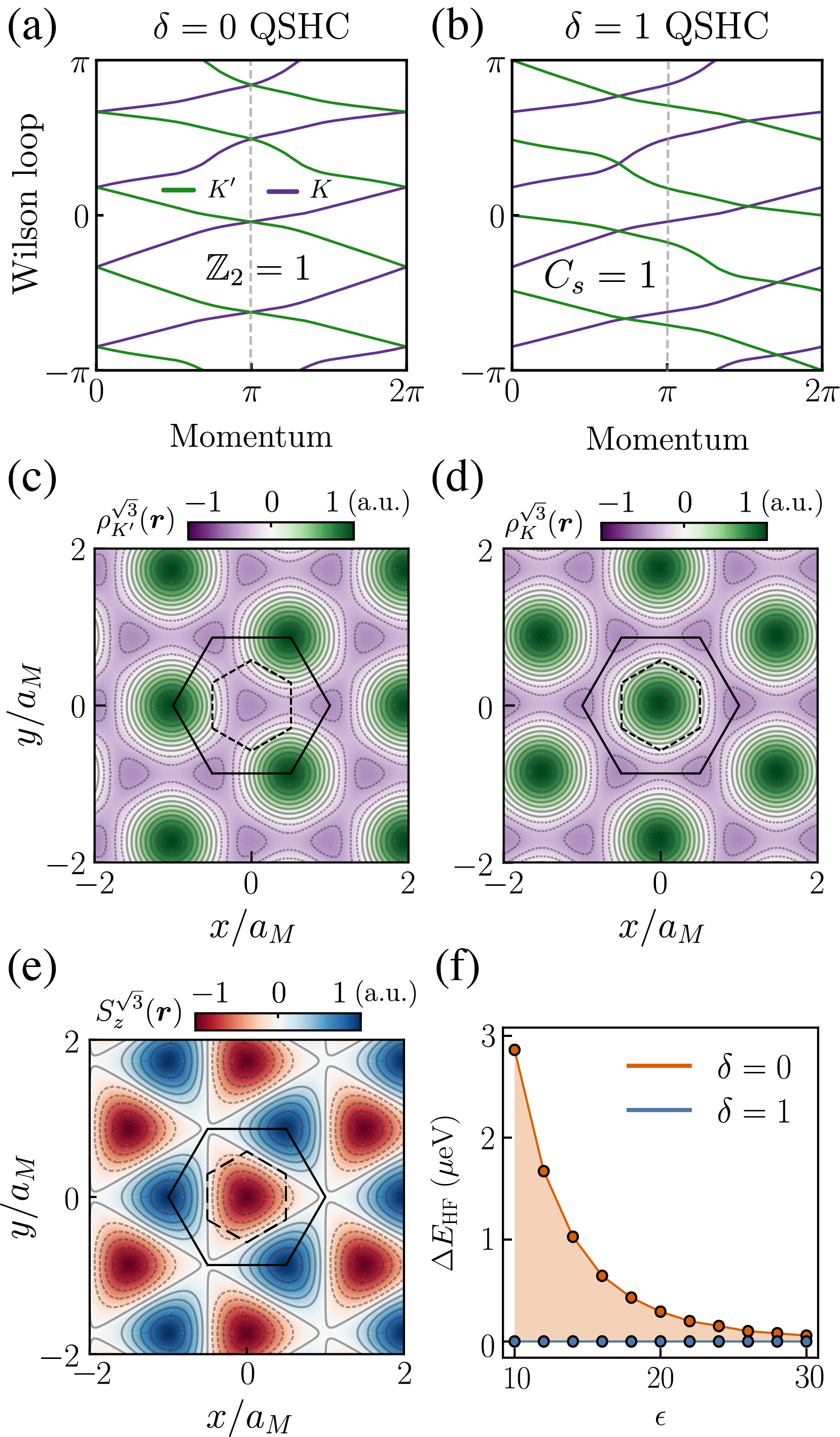}
    \caption{(a,b) Wilson-loop eigenphases of the occupied bands for the $\delta = 0$ and $\delta = 1$ QSHC, respectively, yielding different topological indicators $\mathbb{Z}_2 = 1$ and $C_s = 1$. (c,d) Valley-resolved density modulations $\rho^{\sqrt{3}}_{\tau}(\boldsymbol{r})$ ($\tau = K', K$) as the Fourier transformation of $\rho_\tau(\boldsymbol{Q})$ within the $\sqrt{3}\times\sqrt{3}$ cell for the $\delta = 1$ QSHC.
    (e) The corresponding antiferromagnetic spin texture $S^{\sqrt{3}}_z(\boldsymbol{r})$ of the $\delta = 1$ QSHC. The dashed/solid black hexagon is the primitive cell/ supercell.
    (f) Hartree--Fock energy difference per moiré cell as a function of the dielectric constant $\epsilon$, showing $\delta = 1$ QSHC is slightly favored over $\delta = 0$ QSHC.
    }
    \label{fig:miscellaneous_QSHC}
\end{figure}

\emph{$\mathbb{Z}_3$ structure of the QSHC.}---%
The interplay among translation-symmetry breaking, time-reversal symmetry, and intervalley coupling minimizes the energy of QSHC states into an intrinsic $\mathbb{Z}_3$ structure, making them distinct from a stack of two decoupled QAHCs with opposite Chern numbers.
We now elucidate the origin of this $\mathbb{Z}_3$ structure.

The $\sqrt{3}\times\sqrt{3}$ order gives three copies of crystals related by a moir\'e translation in each valley. Combining the copies of both valleys together, we obtain $3\times3=9$ QSHCs. The nine states are classified into a $\mathbb{Z}_3$ structure by the relative-shift between the two valley crystals,
extracted from the relative phase of their valley-resolved
charge-density-wave components,
\begin{equation}
\rho_\tau(\boldsymbol{Q})
=
\frac{1}{N_{}}\sum_{\boldsymbol{k}\in \text{mBZ}}\lambda_\tau(\boldsymbol{k},\boldsymbol{Q})
\langle
c^\dagger_{\tau,\boldsymbol{k}+\boldsymbol{Q}}
c_{\tau,\boldsymbol{k}}
\rangle,
\quad \tau=K,K',
\end{equation}
where $\lambda_\tau(\boldsymbol{k},\boldsymbol{Q}) = \langle u_{\tau,\boldsymbol{k+Q}}|u_{\tau,\boldsymbol{k}}\rangle$ is the form factor of Bloch states. For the three crystal types, the relative phases are approximated into the $\mathbb{Z}_3$ indices:
\begin{equation}
\frac{\rho_{K'}(\boldsymbol{Q})}{\rho_K(\boldsymbol{Q})}
\approx \omega^\delta,
\qquad
\omega=e^{2\pi i/3},
\qquad
\delta=0,\pm1.
\end{equation}
The index $\delta$ indicates the relative displacement of
the $K'$-valley crystal with respect to the $K$-valley crystal. A
translation by one moiré primitive lattice vector
$\boldsymbol{a}_m$ changes the phase of
$\rho_{K'}(\boldsymbol{Q})$ by
$\boldsymbol{Q}\cdot\boldsymbol{a}_m=2\pi/3$, thereby cycling among the
three sectors.

The structure of this family is most transparent in the
valley-decoupled limit. In this limit, the relative valley translation
$t_s$, which translates one valley by a moiré primitive lattice vector
$\boldsymbol{a}_m$ while leaving the other valley fixed, is an exact
symmetry of the Hamiltonian. It acts on the sector index as
$\delta\to\delta-1$, rendering the three sectors exactly degenerate. When we restore the intervalley coupling, under $t_s$, the intervalley density-density interaction acquires a nontrivial phase
$\omega$, and the relative-translation symmetry is therefore broken. Consequently, the
$\delta=0$ and $\delta=\pm1$ sectors are no longer related by an exact
symmetry, and their degeneracy is lifted (See SM \cite{supplemental} for the detailed discussion). 
Time-reversal symmetry
$\mathcal{T}$, by contrast, remains exact in the presence of the
intervalley interaction and acts on the sector index as $\mathcal{T}:\delta\longmapsto-\delta$. The $\delta=0$ sector is therefore a singlet invariant under $\mathcal{T}$,
whereas the $\delta=\pm1$ sectors individually break time-reversal
symmetry and are exchanged by $\mathcal{T}$, forming an exactly
degenerate doublet.

We thus identify the $\delta=0$ sector as a time-reversal-invariant
QSHC characterized by a $\mathbb{Z}_2$ topological index. The
$\delta=\pm1$ sectors, on the other hand, are 
QSHCs that spontaneously break TRS while preserving $U(1)_v$ and are characterized by the integer
spin-Chern number $C_s\in\mathbb{Z}$. The distinction is visible in the eigenphases of the Wilson loop of the occupied bands, as shown in Fig.~\ref{fig:miscellaneous_QSHC}(a,b). In the $\delta = 0$ QSHC, the eigenphases of the two valleys
cross at time-reversal-invariant momenta (TRIM), whereas in the $\delta = 1$ QSHC
they avoid each other at TRIM. This directly reflects the protecting symmetry:
the topology of the $\delta = 0$ state rests on $\mathcal{T}$ and is
described by a nontrivial $\mathbb{Z}_2$ index, while the $\delta = 1$
state carries $|C_s| = 1$, obtained from the winding of the Wilson-loop eigenphases.
Fig.~\ref{fig:miscellaneous_QSHC}(c,d) compares the $\boldsymbol{Q}$-component of charge distribution $\rho^{\sqrt{3}}_{\tau}(r)$ of two valleys in $\delta = 1$ QSHC. For $\delta = 0$
the charge distributions of the two valleys coincide; for $\delta = 1$ they
are staggered by a moiré lattice vector, so the two modulations superpose
less coherently. Consistent with
its broken TRS, the $\delta = 1$ QSHC develops a weak staggered spin texture, shown in
Fig.~\ref{fig:miscellaneous_QSHC}(e), whereas the $\delta = 0$ QSHC remains nonmagnetic.

Energetically, the $\delta = \pm 1$ branches are slightly more favorable than $\delta = 0$, as shown in Fig.~\ref{fig:miscellaneous_QSHC}(f). The energy difference is extremely small (on the order of $10^{-3}$meV), but persists within a large range of system parameters.
The near-degeneracy is lifted primarily by the intervalley Hartree
channel: for $\delta=0$, the density modulations in
the two valleys are aligned in phase, maximizing their intervalley
Coulomb repulsion. For $\delta=\pm1$, the relative translation staggers
the two density modulations and thereby lowers the Hartree energy (see SM~\cite{supplemental} for details). The $\delta = 0$ QSHC
could instead be stabilized by suppressing the intervalley Coulomb
repulsion, by an attractive intervalley channel such as intervalley
phonons, or by a weak TRS-pinning potential; we leave these directions
for future work.

This analysis shows that, in general, QSHCs cannot be described as a trivial stack of two independent QAHCs.
The intervalley interaction binds the two valleys,
carrying opposite Chern numbers $C_K = -C_{K'} = 1$, into a single
spin-Hall crystal, and selects the relative valley alignment, which in
turn dictates the protecting symmetry of the topology: $\mathcal{T}$
for $\delta = 0$ and $U(1)_v$ for $\delta = \pm 1$. The approximate
$\mathbb{Z}_3$ structure organizes the three sectors into the QSHC
family.

\emph{QSHC families.}---%
The $\mathbb Z_3$ structure above provides a prototypical example of the general
classification of QSHCs. Let us assume that each valley forms the same
commensurate crystal, increasing the periodicity of the system from the moiré lattice $L$ to a superlattice
$L_s$ containing $N$ moiré cells. To label the resulting configurations,
we fix the $K$-valley crystal and displace the $K'$-valley crystal by a
moiré lattice vector $\boldsymbol a_\delta\in L$. Because the crystal is
periodic under $L_s$, displacements differing by a superlattice vector
produce the same configuration. The $N$ distinct relative-shift sectors are
therefore labeled by $\delta\in D\equiv L/L_s$, where $\delta$ has a one-to-one correspondence to $\boldsymbol a_\delta$, with the choice $\boldsymbol a_\delta=-\boldsymbol a_{\delta^\prime}$ mod $L_s$ if and only if $\delta=-\delta^\prime$, and `$/$' denotes the group quotient. 

\begin{table}[t]
\caption{QSHC families for representative crystal patterns. $D = L/L_s$
labels the relative-shift sectors; singlets (s) map to themselves under
$\mathcal{T}$ and doublets (d) are exchanged in exactly degenerate pairs. The
last column lists the topological index of each sector type with its
protecting symmetry: $\mathbb{Z}_2$ for the $\mathcal{T}$-invariant
$\delta = 0$ crystal, $C_s$ ($U(1)_v$) for doublet members, and
$\mathbb{Z}_2$ under the magnetic translation
$\mathcal{T}' = T(\boldsymbol{a}_\delta)\mathcal{T}$ for the
antiferromagnetic singlets of even-index superlattices.}
\label{tab:family}
\begin{ruledtabular}
\begin{tabular}{lccc}
crystal pattern & $D$ & sectors & top. index (symmetry) \\
\hline
$\sqrt{3}\times\sqrt{3}$ & $\mathbb{Z}_3$ & $1\,\mathrm{s} + 1\,\mathrm{d}$ &
$\mathbb{Z}_2\,(\mathcal{T})$;\ $C_s\,(U(1)_v)$ \\
period-2 stripe & $\mathbb{Z}_2$ & $2\,\mathrm{s}$ &
$\mathbb{Z}_2\,(\mathcal{T})$;\ $\mathbb{Z}_2\,(\mathcal{T}')$ \\
$2\times 2$ & $\mathbb{Z}_2\oplus\mathbb{Z}_2$ & $4\,\mathrm{s}$ &
$\mathbb{Z}_2\,(\mathcal{T})$;\ $\mathbb{Z}_2\,(\mathcal{T}')$ \\
$\sqrt{7}\times\sqrt{7}$ & $\mathbb{Z}_7$ & $1\,\mathrm{s} + 3\,\mathrm{d}$ &
$\mathbb{Z}_2\,(\mathcal{T})$;\ $C_s\,(U(1)_v)$ \\
$2\sqrt3\times2\sqrt3$
      & $\mathbb Z_2\oplus\mathbb Z_6$
      & $4\mathrm{s} + 4\mathrm{d}$ &$\mathbb{Z}_2\,(\mathcal{T})$;\ $\mathbb{Z}_2\,(\mathcal{T}')$;\ $C_s\,(U(1)_v)$\\
\end{tabular}
\end{ruledtabular}
\end{table}

In the valley-decoupled limit, independent translations of the two valley
crystals make all $N$ sectors exactly degenerate. Intervalley interactions
lift the degeneracy between inequivalent sectors but preserve the symmetry protecting the topology within each sector. They may
therefore energetically select either an individual self-conjugate sector or
a time-reversal-related pair (see SM \cite{supplemental} for the detailed discussion about interaction selection).

Since $\mathcal{T}$ exchanges the two valleys, it reverses their relative
displacement, i.e., $
    \mathcal T:
    \boldsymbol a_\delta\longmapsto-\boldsymbol a_\delta,
    \delta\longmapsto-\delta
$. If $\delta\neq-\delta$, the two sectors form an exactly degenerate
time-reversal pair. If $\delta=-\delta$, equivalently, $2\boldsymbol a_\delta\in L_s$, the sector is self-conjugate. The $\delta=0$ sector with no relative translation preserves ordinary
$\mathcal{T}$. A nonzero self-conjugate sector is generally invariant only
after time reversal is followed by a lattice translation of the entire system and can therefore
preserve the combined antiunitary symmetry $\widetilde{\mathcal T} =T(\boldsymbol a_\delta)\mathcal T$.
These possibilities are illustrated by the
$\sqrt{3}\times\sqrt{3}$ and $2\times2$ crystals in the following. For the former,
$D\simeq\mathbb Z_3$: only $\delta=0$ is self-conjugate, while the two
nonzero sectors ($\delta=\pm 1$) are exchanged by $\mathcal T$. For a $2\times2$ crystal,
$D\simeq\mathbb Z_2\oplus\mathbb Z_2$, with relative displacements
$\boldsymbol a_\delta=0,\boldsymbol a_1,\boldsymbol a_2$, and
$\boldsymbol a_1+\boldsymbol a_2$. All four satisfy
$2\boldsymbol a_\delta\in L_s$ and are therefore self-conjugate. The zero
sector may preserve ordinary $\mathcal T$, whereas each nonzero sector may
preserve the corresponding combined antiunitary symmetry
$\widetilde{\mathcal T}_\delta
=T(\boldsymbol a_\delta)\mathcal T$.

The classification thus also identifies the symmetry
available to protect the topology. The $\delta=0$ sector is classified by
the usual $\mathbb Z_2$ invariant protected by ordinary $\mathcal T$.
Individual members of a $\mathcal T$-related pair break $\mathcal T$ but,
when $U(1)_v$ is preserved, can be characterized by an integer valley/spin
Chern number $C_s$. Nonzero self-conjugate sectors, which occur only for
even-index superlattices, generally break ordinary $\mathcal T$ but can
support a magnetic $\mathbb Z_2$ topology protected by
$\widetilde{\mathcal T}_\delta$. When this invariant is nontrivial, they
realize antiferromagnetic QSH crystals, analogous to antiferromagnetic
topological insulators \cite{AntiferromagneticTopologicalInsulators_Mong_2010}. Their edge modes are protected only on terminations
that preserve the translation entering
$\widetilde{\mathcal T}_\delta$. Table~\ref{tab:family} summarizes
representative patterns.

\emph{Discussion.}---%
In this Letter, we have demonstrated that
fractionally filled tMoTe$_2$ at $\nu = -8/3$ hosts an
interaction-driven QSHC. Unlike the QSHI at integer filling, the QSHC
requires a finite interaction strength and competes with an antiferromagnetic state and the IVC crystal. Within the QSHC phase, the intervalley interaction
further organizes the ground states into an approximate $\mathbb{Z}_3$
family, comprising a time-reversal-invariant $\mathbb{Z}_2$ member and
a time-reversal-breaking doublet.
We have also provided a general classification of QSHCs.

Several experimental signatures follow: the QSHC combines a quantized
spin Hall response with the $\sqrt{3}\times\sqrt{3}$ charge order at $\nu = -8/3$,
detectable by transport and by either scanning tunneling
microscopy or scanning single electron transistor (SET) measurements \cite{li2021imaging,li2024wigner,ImagingSubMoirePotential_Klein_2026,MicroscopicSignaturesTopology_Thompson_2025,Tsui_2024}. Furthermore, the transition into the IVC
crystal could be tracked through the loss of the helical edge
conduction \cite{konig2008quantum,roth2009nonlocal,jin2026observationmottquantumspin}. Owing to distinct magnetism, the QSHC, AFM$_z$, and IVC crystals could be distinguished via magnetic circular dichroism or local magnetometry \cite{ObservationFerromagneticPhase_An_2025,sun2026twist}.

Finally, our results suggest several future directions. The near-degeneracy of the
$\mathbb{Z}_3$ family invites a study of its collective excitations
and domain physics, including the possibility of thermal or
disorder-driven transitions among the sectors and the associated breaking or restoration of time-reversal symmetry. More broadly, the
interplay between the crystal order and fractionalization remains
open: whether the QSHC at $\nu = -8/3$ gives way to a fractional
quantum spin Hall state (FQSHE) \cite{FractionalTI_Levin_2009,EvidenceFractionalQuantum_Kang_2024,Neupert_2015,PhysRevB.90.245401,NonAbelianSHI_Abouelkomsan_2025,PhysRevB.110.045114,RegardingExistenceAbelian_Kwan_2026,roy2014band,bvrb-z4hj} at nearby fillings, and how the
$\mathbb{Z}_3$ structure evolves in that regime, are natural questions worth exploring.

\emph{Acknowledgements.}---%
We thank Chong Wang for helpful discussions. X.S. is partly supported by the Tsinghua scholarship for overseas graduate studies. This work was supported by the Swedish Research Council (VR, grant 2024-04567), the Wallenberg Scholars program of the Knut and Alice Wallenberg Foundation (2023.0256), and the G\"oran Gustafsson Foundation for Research in Natural Sciences and Medicine.

\bibliography{reference.bib}
\section{End Matter}


\label{sec:EM-sectors}

\paragraph{QSHC classification derivation.--}
In this section, we elaborate on the classification of the relative-shift sectors of QSHC families for general
commensurate superlattices.

Let $L$ be the
moir\'e Bravais lattice and $L_s\subset L$ the superlattice.  Their primitive
vectors obey $(\boldsymbol A_1\ \boldsymbol A_2)=
(\boldsymbol a_1\ \boldsymbol a_2)\mathsf M$, where
$\boldsymbol a_i$ and $\boldsymbol A_i$ belong to $L$ and $L_s$,
respectively, and $N=|\det\mathsf M|$.  For a
$\sqrt3\times\sqrt3$ crystal,
$\mathsf M=\bigl(\begin{smallmatrix}1&-1\\1&2\end{smallmatrix}\bigr)$ and
$N=3$.
The ordering vectors $\boldsymbol{g}$ belong to $L_s^*/L^*$, the group quotient between the reciprocal lattices of the super and moir\'e lattices.  Fixing the crystal in one
valley and translating that in the other by $a_\delta$, inequivalent relative shifts are
labeled by $\delta\in D\equiv L/L_s$.  For a representative
$\boldsymbol a_\delta\in L$, the character
$\chi_\delta(\boldsymbol g)=
e^{i\boldsymbol g\cdot\boldsymbol a_\delta}$ is well defined and
distinguishes all $N$ shifts.  Indeed, it is unchanged under
$\boldsymbol a_\delta\mapsto\boldsymbol a_\delta+\boldsymbol s$ with
$\boldsymbol s\in L_s$, since
$\boldsymbol g\cdot\boldsymbol s\in2\pi\mathbb Z$, and under
$\boldsymbol g\mapsto\boldsymbol g+\boldsymbol G$ with
$\boldsymbol G\in L^*$, since
$\boldsymbol G\cdot\boldsymbol a_\delta\in2\pi\mathbb Z$.  The
$N$ elements of $L_s^*/L^*$ form the character group dual to $D$; the
pairing in $\chi_\delta(\boldsymbol g)$ is nondegenerate, so no two relative
shifts have the same set of characters.  In the
$\sqrt3\times\sqrt3$ case, $D\simeq\mathbb Z_3$.  A suitable choice of primitive moir\'e
translation $\boldsymbol a_m$ and ordering vector $\boldsymbol Q$ satisfies
$e^{i\boldsymbol Q\cdot\boldsymbol a_m}=\omega$ with $\boldsymbol Q \in L_s^*$, so the three sectors carry
$(1,\omega,\omega^2)$, with $\omega=e^{2\pi i/3}$, namely $\chi_\delta(\boldsymbol Q)=\omega^\delta$. The $\delta=0$ sector is self-conjugate under time reversal and thus forms a singlet, whereas the $\delta=\pm 1$ sectors are interchanged by time reversal and form a doublet. The character $\chi_\delta(\boldsymbol{g})$ occurs naturally in the intervalley Hartree energy, see SM~\cite{supplemental}.

In general superlattices, self-conjugate singlets satisfy
$2\boldsymbol a_\delta\in L_s$; all remaining sectors occur in pairs exchanged by time reversal. For the singlets, we have $\delta=-\delta$.
The $\delta=0$ sector with no relative translation preserves ordinary
$\mathcal{T}$. A nonzero self-conjugate sector is generally invariant only
after time reversal is followed by a lattice translation of the entire system and can therefore
preserve the combined antiunitary symmetry $\widetilde{\mathcal T} =T(\boldsymbol a_\delta)\mathcal T$.
On the other hand, any pair $(\delta,-\delta)$ with $\delta\neq -\delta$ constitute a time-reversal-conjugate doublet.

We can count the numbers of singlets and doublets by the Smith normal form of $\mathsf M$. The integer matrix $\mathsf M$ can be decomposed as $\mathsf M=USV$, where $U,V$ are invertible integer matrices with determinant $\pm1$, and $S=\text{diag}(d_1,d_2)$ with integer $d_1,d_2$. Here, $d_1=\text{gcd}(\text{entries of }\mathsf M)$, $d_1$ divides $d_2$ and $d_1d_2=N$. From $(\boldsymbol A_1\ \boldsymbol A_2)=
(\boldsymbol a_1\ \boldsymbol a_2)\mathsf M$, we see $U,V$ are linear transformations of $\boldsymbol a_i,\boldsymbol A_i$ into parallel directions, namely $\boldsymbol A^\prime_i=d_i\boldsymbol a^\prime_i$ for the new basis $\boldsymbol a^\prime_i,\boldsymbol A^\prime_i$. Therefore, we obtain $D\simeq\mathbb Z_{d_1}\oplus\mathbb Z_{d_2}$. We can also extract that the number of self-conjugate sectors is
  $N_s=\gcd(2,d_1)\,\gcd(2,d_2)$. In other words, if both $d_1,d_2$ are even, there are 4 singlets; if one is even and the other is odd, there are 2 singlets; if both are odd, then the zero-shift sector is the only singlet.
The remaining $(N-N_s)$ sectors form $(N-N_s)/2$ time-reversal-exchanged pairs. We work out a few representative cases, as listed in Table~\ref{tab:family}. Note that the index $N$ alone does not determine the sector structure. For example, a
$1\times4$ stripe has $D\simeq\mathbb Z_4$ and consists of two singlets and
one doublet, whereas a $2\times2$ crystal has
$D\simeq\mathbb Z_2\oplus\mathbb Z_2$ and consists of four singlets, although
both have $N=4$.

\newpage
\clearpage
\onecolumngrid
\vspace{1cm}
\begin{center}
    {\bf\large Supplemental Materials for ``Quantum spin Hall crystals at fractional filling of twisted MoTe$_2$''}
\end{center}
\tableofsmcontents

\smsection{Continuum model}

The single-particle continuum Hamiltonian for twisted bilayer
MoTe$_2$ \cite{TopologicalInsulatorsTwisted_Wu_2019} in valley $\tau=\pm1$ is
\begin{equation}
\label{eq:continuum-hamiltonian}
H_{0,\tau} =
\begin{pmatrix}
-\dfrac{\hbar^2
\left(\boldsymbol{k}-\tau\boldsymbol{\kappa}_{+}\right)^2}
{2m^*}
+\Delta_{\mathfrak b}(\boldsymbol r)+V_z/2
&
\Delta_T(\boldsymbol r)
\\[6pt]
\Delta_T^\dagger(\boldsymbol r)
&
-\dfrac{\hbar^2
\left(\boldsymbol{k}-\tau\boldsymbol{\kappa}_{-}\right)^2}
{2m^*}
+\Delta_{\mathfrak t}(\boldsymbol r)-V_z/2
\end{pmatrix},
\end{equation}
where $\boldsymbol{\kappa}_{\pm}$ denote the valley-dependent
momentum offsets in the two layers and $V_z$ is the displacement field. The intralayer moiré potentials
$\Delta_{\mathfrak t,\mathfrak b}$ and the interlayer tunneling
amplitude $\Delta_T$ are given by
\begin{align}
\Delta_{\mathfrak t,\mathfrak b}(\boldsymbol r)
&=
2V\sum_{j=1,3,5}
\cos\left(
\boldsymbol G_j\cdot\boldsymbol r+\ell\psi
\right),
\label{eq:moire-potential}
\\
\Delta_T(\boldsymbol r)
&=
w\left[
1+
e^{-i\tau\boldsymbol G_2\cdot\boldsymbol r}
+
e^{-i\tau\boldsymbol G_3\cdot\boldsymbol r}
\right].
\label{eq:interlayer-tunneling}
\end{align}
Here, $\ell=\pm1$ labels the two layers, and
$\boldsymbol G_j$ ($j=1,\ldots,6$) are the moiré reciprocal-lattice
vectors. We use continuum-model parameters obtained by fitting to
first-principles calculations \cite{FractionalChernInsulator_Wang_2024}:
\begin{equation}
V=20.8~\mathrm{meV},
\qquad
\psi=107.7^\circ,
\qquad
w=-23.8~\mathrm{meV}.
\end{equation}
The monolayer lattice constant is taken to be
$a_0=3.52~\text{\AA}$, and the effective mass is
$m^*=0.62m_e$, where $m_e$ is the bare electron mass.

Including Coulomb interactions, the projected many-body Hamiltonian
takes the form
\begin{equation}
\label{eq:interacting-hamiltonian}
H
=
H_0+
\frac{1}{2A}
\sum_{\boldsymbol q}
V(q)
:
\hat{\rho}(\boldsymbol q)
\hat{\rho}(-\boldsymbol q)
: ,
\end{equation}
where $A$ is the area of the two-dimensional system,
$q=|\boldsymbol q|$ is the magnitude of the transferred momentum,
and $\hat{\rho}(\boldsymbol q)$ is the density operator projected
onto the active bands. We model the screened Coulomb interaction as
\begin{equation}
\label{eq:screened-interaction}
V(q)
=
\eta_{\mathrm{RK}}(q)
\frac{e^2\tanh(qd)}
{2\epsilon\epsilon_0 q},
\qquad
\eta_{\mathrm{RK}}(q)
=
\frac{1}{1+\ell_{\mathrm{RK}}q}.
\end{equation}
Here, the factor $\tanh(qd)$ accounts for screening by metallic gates,
while $\eta_{\mathrm{RK}}(q)$ is the Rytova--Keldysh correction
associated with the finite in-plane polarizability and thickness of
the material \cite{Rytova1967,Keldysh1979,RegardingExistenceAbelian_Kwan_2026}. We take the Rytova--Keldysh screening length to be
$\ell_{\mathrm{RK}}=2.7~\mathrm{nm}$ and the gate-to-sample distance
to be $d=10~\mathrm{nm}$. Finally, $e$ is the elementary charge,
$\epsilon_0$ is the vacuum permittivity, and $\epsilon$ is the
relative dielectric constant.

\smsection{Energy analysis}
\label{sec:SM-sectors}

In this section, we analyze the energy degeneracy and splitting among the QSHC families.

We work in the active-band subspace of each valley and suppress the band
indices for notational simplicity. A state preserving $L_s$ translational symmetry and the valley
charge $U(1)_v$ is characterized by the one-body density matrix
\begin{equation}
  P_\tau(\boldsymbol k,\boldsymbol k')
  =\left\langle
  c^\dagger_{\tau,\boldsymbol k'}c_{\tau,\boldsymbol k}
  \right\rangle,
  \qquad
  \label{eq:density-matrix-SM}
\end{equation}
with $P_\tau(\boldsymbol k,\boldsymbol k')\neq0
  \ \text{only if}\ 
  \boldsymbol k'-\boldsymbol k\in L_s^*$. The
momentum-off-diagonal components describe the crystal order. In the same
convention, the projected density operator and its expectation value are
\begin{equation}
  \begin{aligned}
    \hat{\bar\rho}_\tau(\boldsymbol q)
    &=\sum_{\boldsymbol k}
      \lambda_\tau(\boldsymbol k,\boldsymbol q)
      c^\dagger_{\tau,\boldsymbol k+\boldsymbol q}
      c_{\tau,\boldsymbol k},\\
    \rho_\tau(\boldsymbol q)
    &\equiv\langle\hat{\bar\rho}_\tau(\boldsymbol q)\rangle
      =\sum_{\boldsymbol k}
      \lambda_\tau(\boldsymbol k,\boldsymbol q)
      P_\tau(\boldsymbol k,\boldsymbol k+\boldsymbol q),
  \end{aligned}
  \label{eq:density-harmonics-SM}
\end{equation}
with the form factor $\lambda_\tau(\boldsymbol k,\boldsymbol q)
    =\langle u_{\tau,\boldsymbol k+\boldsymbol q}
      |u_{\tau,\boldsymbol k}\rangle$ satisfying 
$\lambda_\tau(\boldsymbol k,\boldsymbol q)^*
=\lambda_\tau(\boldsymbol k+\boldsymbol q,-\boldsymbol q)$. In a
time-reversal gauge,
$\lambda_{K'}(\boldsymbol k,\boldsymbol q)
=\lambda_K(-\boldsymbol k,-\boldsymbol q)^*$.

For a density-density interaction, the Hartree--Fock energy under the $U(1)_v$ symmetry is
\begin{align}
  E[P]
  &=\sum_{\tau,\boldsymbol k}
    \varepsilon_\tau(\boldsymbol k)
    P_\tau(\boldsymbol k,\boldsymbol k)
  +\frac{1}{2A}\sum_{\boldsymbol q\in L_s^*}V(q)
    \sum_{\tau,\tau'}
    \rho_\tau(\boldsymbol q)\rho_{\tau'}(-\boldsymbol q)
  \label{eq:EHF-SM}\\
  &\quad
  -\frac{1}{2A}\sum_{\tau,\boldsymbol q}V(q)
    \sum_{\boldsymbol k,\boldsymbol k'}
    \lambda_\tau(\boldsymbol k,\boldsymbol q)
    \lambda_\tau(\boldsymbol k',\boldsymbol q)^*
    P_\tau(\boldsymbol k'+\boldsymbol q,
           \boldsymbol k+\boldsymbol q)
    P_\tau(\boldsymbol k,\boldsymbol k').
  \nonumber
\end{align}
The uniform Hartree contribution is understood to be canceled by the
neutralizing background. Due to the $U(1)_v$ symmetry, there is no
intervalley coherence. Therefore, the Fock term is purely intravalley, and the
only intervalley contribution is the Hartree cross term with
$\tau\neq\tau'$.

Translating the $K'$ crystal by $\boldsymbol a\in L$ while leaving the $K$
crystal fixed acts as
\begin{equation}
  t_s(\boldsymbol a):\quad
  P_{K'}(\boldsymbol k,\boldsymbol k')
  \longmapsto
  e^{i(\boldsymbol k'-\boldsymbol k)\cdot\boldsymbol a}
  P_{K'}(\boldsymbol k,\boldsymbol k'),
  \label{eq:ts-SM}
\end{equation}
and hence
$\rho_{K'}(\boldsymbol q)\mapsto
e^{i\boldsymbol q\cdot\boldsymbol a}\rho_{K'}(\boldsymbol q)$.
For an $L_s$-periodic crystal, this transformation depends on
$\boldsymbol a$ only through its equivalence class in $D$ and maps one
relative-shift sector to another. Under Eq.~\eqref{eq:ts-SM}, the kinetic term is unchanged because it contains
only momentum-diagonal elements. The phases also cancel pairwise in the
intravalley Hartree and Fock terms. By contrast, the intervalley Hartree term
$\rho_K(\boldsymbol q)\rho_{K'}(-\boldsymbol q)$ acquires an uncompensated
phase $e^{-i\boldsymbol q\cdot\boldsymbol a}$, which is nontrivial for
$\boldsymbol q\in L_s^*\setminus L^*$. We define the valley-decoupled limit
by removing this intervalley Hartree cross term while retaining the
single-particle and intravalley interaction terms. In this limit, any
self-consistent solution generates an orbit of $N$ exactly degenerate
solutions labeled by $\delta\in D$. Restoring the intervalley interaction
breaks the independent relative-translation symmetry and can split
inequivalent sectors.

Time reversal exchanges the two valleys and reverses their relative displacement, i.e.,  $\mathcal T:\quad \boldsymbol{a}_\delta\longmapsto-\boldsymbol{a}_\delta$. 
Under time reversal, the Fourier harmonics transform as
\begin{equation}
  \rho_{K'}^{\mathcal T}(\boldsymbol g)
  =\rho_K(-\boldsymbol g)^*
  =\rho_K(\boldsymbol g).
  \label{eq:TR-density-SM}
\end{equation}
The second equality follows from the reality of the real-space density.
Time reversal remains exact when the intervalley interaction is restored.
The sectors therefore organize into self-conjugate sectors satisfying
$2\delta=0$ and exchanged pairs satisfying $\delta\neq-\delta$, as we have discussed in the main text. Combining Eq.~\eqref{eq:EHF-SM} and \eqref{eq:TR-density-SM}, we see the two
members of every exchanged pair have exactly equal energies,
\begin{equation}
  E_\delta=E_{-\delta}.
  \label{eq:TR-energy-SM}
\end{equation}

As an illustration, we consider a $2\times2$ crystal. Period doubling in both
directions, $\mathsf M=2\mathbb I$, gives $N=4$ and
$D\simeq\mathbb Z_2\oplus\mathbb Z_2$. All four sectors are self-conjugate,
and no time-reversal doublet occurs. The three ordering vectors satisfy
$\boldsymbol M_3=\boldsymbol M_1+\boldsymbol M_2$ modulo $L^*$, and their
characters, defined in the End Matter of the main text, form the sign triplets
\begin{equation}
  \begin{split}
  \bigl(\chi_\delta(\boldsymbol M_1),
        \chi_\delta(\boldsymbol M_2),
        \chi_\delta(\boldsymbol M_3)\bigr)
  ={}( +,+,+),\ (-,+,-),\
     (+,-,-),\ (-,-,+).
  \end{split}
  \label{eq:2x2-characters}
\end{equation}
Assuming that a self-conjugate solution preserves the corresponding
$\widetilde{\mathcal T}_\delta$, its valley-resolved density harmonics obey
\begin{equation}
  \rho_{K'}(\boldsymbol M_i)
  =\chi_\delta(\boldsymbol M_i)\rho_K(\boldsymbol M_i).
  \label{eq:2x2-valley-relation}
\end{equation}
Defining the charge and valley/spin harmonics as
\begin{align}
  \rho_c(\boldsymbol M_i)
  &=\rho_K(\boldsymbol M_i)+\rho_{K'}(\boldsymbol M_i),\\
  \rho_v(\boldsymbol M_i)
  &=\rho_K(\boldsymbol M_i)-\rho_{K'}(\boldsymbol M_i),
\end{align}
we obtain
\begin{align}
  \rho_c(\boldsymbol M_i)
  &=\bigl[1+\chi_\delta(\boldsymbol M_i)\bigr]
    \rho_K(\boldsymbol M_i),\\
  \rho_v(\boldsymbol M_i)
  &=\bigl[1-\chi_\delta(\boldsymbol M_i)\bigr]
    \rho_K(\boldsymbol M_i).
  \label{eq:2x2-channel-selection}
\end{align}
Thus, a nonzero harmonic with $\chi_\delta=+1$ belongs to the charge
channel, whereas one with $\chi_\delta=-1$ belongs to the valley/spin
channel. If the reference $2\times2$ crystal is a $C_3$-symmetric triple-$M$
state, the zero-shift sector contains three charge harmonics and can preserve
both $C_3$ and ordinary time reversal. Each nonzero sector then contains one
charge harmonic and two valley/spin harmonics, corresponding to a charge
stripe accompanied by a collinear $s_z$ spin-density wave. For a
$C_3$-symmetric Hamiltonian, the three nonzero sectors are related by
rotation and therefore have equal energies. If
$\widetilde{\mathcal T}_\delta$ is further broken, charge and valley/spin
components may coexist at the same ordering vector.

We finally discuss the energetic splitting between different QSHC sectors.
We treat the intervalley Hartree interaction as a perturbation to the
valley-decoupled QSHC manifold by introducing a parameter
$\eta$,
\begin{equation}
  E_\eta[P]
  =E_{\mathrm{dec}}[P]+\eta E_{\mathrm H}^{KK'}[P],
  \qquad
  E_{\mathrm H}^{KK'}[P]
  =\frac{1}{A}\sum_{\boldsymbol q}V(q)
   \rho_K(\boldsymbol q)\rho_{K'}(-\boldsymbol q).
  \label{eq:eta-expansion-SM}
\end{equation}
At $\eta=0$, the $N$ degenerate zeroth-order solutions are related by
\begin{equation}
  P_\delta^{(0)}
  =t_s(\boldsymbol a_\delta)P_0^{(0)}.
\end{equation}
With the convention in Eq.~\eqref{eq:ts-SM}, their density harmonics satisfy
\begin{equation}
  \rho_{K,\delta}^{(0)}(\boldsymbol g)=\rho_0(\boldsymbol g),
  \qquad
  \rho_{K',\delta}^{(0)}(\boldsymbol g)
  =e^{i\boldsymbol g\cdot\boldsymbol a_\delta}
   \rho_0(\boldsymbol g).
  \label{eq:zeroth-order-density-SM}
\end{equation}
The first-order correction is the intervalley Hartree energy evaluated on
the unperturbed QSHC,
\begin{align}
  E_\delta^{(1)}
  &=E_{\mathrm H}^{KK'}[P_\delta^{(0)}]\nonumber\\
  &=E_{\mathrm{const}}^{(1)}
  +\frac{1}{A}
   \sum_{\boldsymbol g\in L_s^*\setminus L^*}
   V(g)|\rho_0(\boldsymbol g)|^2
   \cos(\boldsymbol g\cdot\boldsymbol a_\delta),
  \label{eq:lockin-SM}
\end{align}
where $E_{\mathrm{const}}^{(1)}$ contains the sector-independent
contributions. The sector energy therefore has the expansion
\begin{equation}
  E_\delta(\eta)
  =E_{\mathrm{dec}}+\eta E_\delta^{(1)}+O(\eta^2).
  \label{eq:sector-energy-expansion-SM}
\end{equation}
Because $P_\delta^{(0)}$ is a stationary solution of
$E_{\mathrm{dec}}[P]$, the first-order change of the density matrix does not
contribute to the energy at first order. Independent self-consistent
relaxation of the different sectors therefore enters through the
$O(\eta^2)$ and higher-order terms.

Equation~\eqref{eq:lockin-SM} satisfies
$E_\delta^{(1)}=E_{-\delta}^{(1)}$ term by term. More generally,
$E_\delta=E_{-\delta}$ remains exact to all orders because it is protected
by time reversal. The magnitude of the first-order splitting is controlled
by the projected density harmonics $|\rho_0(\boldsymbol g)|^2$, which depend
on both the band form factors and the strength of the crystal order.

For a repulsive interaction, $V(g)>0$, the aligned configuration
$\delta=0$ maximizes the first-order intervalley Hartree energy. For the
$\sqrt3\times\sqrt3$ crystal, the shifted $\delta=\pm1$ sectors are therefore
favored at first order, with
\begin{equation}
  E_{\pm1}^{(1)}-E_0^{(1)}=-\frac{3}{2}J_Q,
  \qquad
  J_Q=\frac{1}{A}\sum_{\boldsymbol g\in Q\text{-star}}
  V(g)|\rho_0(\boldsymbol g)|^2>0.
\end{equation}
For a $C_3$-symmetric $2\times2$ triple-$M$ crystal,
Eq.~\eqref{eq:lockin-SM} reduces to
\begin{equation}
  E_\delta^{(1)}
  =E_{\mathrm{const}}^{(1)}
  +\sum_iJ_i\chi_\delta(\boldsymbol M_i),
  \qquad
  J_i=\frac{V(M_i)}{A}|\rho_0(\boldsymbol M_i)|^2>0.
\end{equation}
When $J_i=J$, this gives
$E_0^{(1)}=E_{\mathrm{const}}^{(1)}+3J$ and
$E_{\delta\neq0}^{(1)}=E_{\mathrm{const}}^{(1)}-J$.

At the physical value $\eta=1$, the higher-order corrections need not be
small. They incorporate the independent relaxation of the density profiles
and modify all components of the self-consistent Hartree--Fock energy, so
they may reduce or even reverse the first-order preference between
inequivalent time-reversal orbits. The final sector ordering is therefore a
quantitative property of the full self-consistent calculation, whereas the
degeneracy within each time-reversal-related pair remains exact.

\smsection{One-band-per-valley Exact diagonalization}
\label{sec:ED}

\begin{figure}
    \centering
    \includegraphics[width=0.9\linewidth]{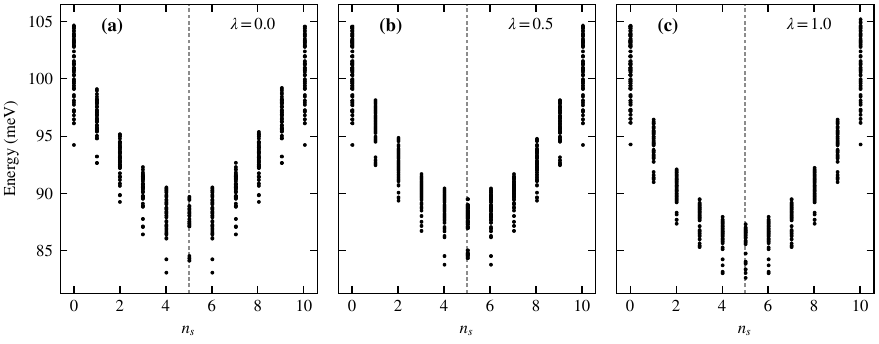}
    \caption{One band per valley exact diagonalization for the second moiré band at $\theta=5.25^\circ$, $V_z=10$meV, and $\epsilon=17$. Here, $\lambda$ changes the intervalley interaction strength, $n_s$ is the occupation number in valley $K$, and $n_s=5$ indicates the symmetric filling $S_z=0$ for the tilted sample $\mathbf{T}_1=(4, 1)$ and $\mathbf{T}_2=(1, 4)$. 
    }
    \label{fig:1BPV}
\end{figure}

In the main text, we noted that the ground states of tMoTe$_2$ at large twist angles are not valley polarized. 
Here, we further support this observation using exact diagonalization within a one-band-per-valley model for the second moiré band at $2/3$ filling. 
We take the intravalley and intervalley interactions to be $v_{\text{intra}}=\eta_{\mathrm{RK}}(q)\frac{e^2\tanh(qd)}{2\epsilon\epsilon_0 q}$ and $v_{\text{inter}}=\lambda \eta_{\mathrm{RK}}(q)\frac{e^2\tanh(qd)}{2\epsilon\epsilon_0 q}$, where $\lambda$ controls the relative strength of the intervalley interactions.
As shown in Fig.~\ref{fig:1BPV}, even when the intervalley interaction is completely switched off ($\lambda=0$), the system does not lie in the fully valley-polarized sector. 
Upon increasing $\lambda$ to $1$, where inter- and intravalley have equal interaction strengths, the ground state instead favors the valley-balanced sector with $S_z=0$. 

We nevertheless observe strong competition from the nearby $S_z=1$ sector. We attribute this near-degeneracy to finite-size effects and the lack of band-mixing effects, since increasing the dielectric constant, and hence reducing the overall interaction strength, further favors the valley-balanced sector. 
For the small system sizes accessible to exact diagonalization, however, the expected crystalline signatures are not yet clearly resolved within the $S_z=0$ sector. A full exact-diagonalization calculation retaining two active bands in each valley, which would
allow band-mixing effects to be assessed, is beyond the scope of the present work. Despite this limitation, the absence of a valley-polarized phase over the interaction range considered supports our focus on time-reversal-related competing orders in the large-twist-angle regime. Our exact-diagonalization results should therefore be interpreted as evidence against complete valley polarization over the parameter range considered, rather than as an independent identification of the QSHC.

\smsection{Self-consistent Hartree--Fock calculations}
\label{sec:SCHF-numerics}

We perform self-consistent Hartree--Fock (SCHF) calculations for twisted
bilayer MoTe$_2$, retaining both valleys and allowing for
translation-symmetry-breaking orders through an enlarged unit cell and its
corresponding crystal Brillouin zone (cBZ).

We use the Bloch-band operators $c_{\tau n,\boldsymbol k}$, density matrix
$P$, and form factors $\lambda$ introduced above, restoring the band index
$n$ where needed. For a translational-symmetry-breaking crystal with real-space lattice $L_s$, let
$\boldsymbol Q_{1,2}$ be primitive vectors of $L_s^*$, the reciprocal lattice of $L_s$. The number of folded
momenta is
\[
  N
  =\frac{|\boldsymbol b_1\times\boldsymbol b_2|}
         {|\boldsymbol Q_1\times\boldsymbol Q_2|},
\]
where $\boldsymbol b_{1,2}$ span the original moir\'e reciprocal lattice and
$N=|L/L_s|$ is the number of moir\'e cells in the crystal unit cell defined
above. Any momentum $\widetilde{\boldsymbol k}$ in the moir\'e Brillouin zone
can be written as
$\widetilde{\boldsymbol k}=\boldsymbol k+\boldsymbol Q$, where
$\boldsymbol k$ lies in the cBZ and $\boldsymbol Q$ belongs to a set of
$N_F$ representatives of $L_s^*/L^*$. We collect the valley, folding-vector,
and band labels into the composite index
$\alpha=(\tau,\boldsymbol Q,n)$.

The quantities introduced above are now regarded as matrices in this
composite space. Within a valley-diagonal block,
$P_{\alpha\beta}(\boldsymbol k)$ is simply the band-resolved component of
$P_\tau(\boldsymbol k+\boldsymbol Q_\alpha,
        \boldsymbol k+\boldsymbol Q_\beta)$, with exactly the convention of
Eq.~\eqref{eq:density-matrix-SM}. The numerical matrix also retains
valley-off-diagonal blocks, allowing intervalley-coherent states. Likewise,
$[\lambda(\boldsymbol k,\boldsymbol q)]_{\beta\alpha}$ denotes the
multiband folded form factor connecting the initial state
$(\boldsymbol k,\alpha)$ to the final state
$(\boldsymbol k+\boldsymbol q,\beta)$, using the definition in
Eq.~\eqref{eq:density-harmonics-SM}. Momenta outside the cBZ are folded back
by changing the folding-vector index. In this representation, the density
harmonic becomes
\[
  \rho(\boldsymbol g)
  =\sum_{\boldsymbol k}
  \mathrm{Tr}\!\left[
    P(\boldsymbol k)
    \lambda(\boldsymbol k,\boldsymbol g)
  \right],
  \qquad \boldsymbol g\in L_s^*.
\]

Using the screened Coulomb interaction $V(q)$ defined in
Eq.~\eqref{eq:screened-interaction}, the Hartree and Fock contributions to the mean-field
Hamiltonian are
\begin{align}
  h_H[P](\boldsymbol k)
  &=\frac{1}{A}\sum_{\boldsymbol g\in L_s^*}V(g)
    \lambda(\boldsymbol k,\boldsymbol g)^\dagger
    \left\{
      \sum_{\boldsymbol k'}
      \mathrm{Tr}\!\left[
        P(\boldsymbol k')
        \lambda(\boldsymbol k',\boldsymbol g)
      \right]
    \right\},
  \label{eq:numerical-Hartree}\\
  h_F[P](\boldsymbol k)
  &=-\frac{1}{A}\sum_{\boldsymbol q}V(q)
    \lambda(\boldsymbol k,\boldsymbol q)^\dagger
    P(\boldsymbol k+\boldsymbol q)
    \lambda(\boldsymbol k,\boldsymbol q).
  \label{eq:numerical-Fock}
\end{align}
Here, $\boldsymbol g$ runs over reciprocal vectors of the crystal
superlattice, whereas $\boldsymbol q$ runs over all momentum transfers
compatible with the momentum grid. The uniform Hartree contribution is
canceled by the neutralizing background. The positive Hartree and negative
Fock signs agree with the energy functional in Eq.~\eqref{eq:EHF-SM}.

The mean-field Hamiltonian is
$h_{\mathrm{HF}}[P]=h_{\mathrm{kin}}+h_H[P]
+h_F[P]$, where $h_{\mathrm{kin}}$ is the projection of $H_0$ onto
the active-band subspace. The energy of the corresponding Slater determinant
is
\begin{equation}
  E[P]
  =\frac{1}{2}\mathrm{Tr}\!\left[
    P
    \bigl(2h_{\mathrm{kin}}+h_H[P]+h_F[P]\bigr)
  \right],
  \label{eq:numerical-HF-energy}
\end{equation}
where the trace includes the cBZ momenta and all internal indices. This
expression is equivalent to Eq.~\eqref{eq:EHF-SM}; the factor of $1/2$
removes the double counting of the interaction energy.

At the required filling, we solve
$[P,h_{\mathrm{HF}}[P]]=0$ together with
$P^2=P$. Starting from an initial density matrix, we
construct $h_{\mathrm{HF}}$, diagonalize it, refill the lowest-energy states,
and update $P$ with damping until convergence. Multiple initial
conditions are used to search for competing self-consistent solutions, including the valley-polarized, IVC, $U(1)_v$ non-polarized, and at least 20 random ansätze that consist of various gapped and gapless initial states. The convergence threshold is set within $|[P,h_{\text{HF}}[P]]|\leq 10^{-4}\text{meV}$ and $\Delta E =  E_{n+1}-E_{n} \leq 10^\text{-5}\text{meV}$ at the iteration step $n$. In
particular, the fully converged sector energies include the higher-order
relaxation effects
rather than only the first-order intervalley Hartree splitting.

We determine the topology of the converged SCHF states using the
Fukui--Hatsugai--Suzuki construction \cite{fukuichern} and Wilson loops. We first describe the
general calculation, which also applies when intervalley-coherent order
breaks $U(1)_v$.

Let $|\Psi_{\boldsymbol k n}\rangle$ denote the occupied SCHF eigenstates in
the combined two-valley basis. For a link
$\boldsymbol k_a\rightarrow\boldsymbol k_b$ on the discretized cBZ, we form
the non-Abelian overlap matrix
\[
  [M_{ab}]_{nn'}
  =\langle\Psi_{\boldsymbol k_a n}
  |\Psi_{\boldsymbol k_b n'}\rangle.
\]
Links crossing the cBZ boundary are closed using the appropriate
reciprocal-lattice sewing transformation in each valley. For numerical
stability, each overlap matrix is replaced by the unitary factor in its
polar decomposition,
\[
  \widetilde M_{ab}
  =\mathcal U[M_{ab}]
  =UV^\dagger,
  \qquad
  M_{ab}=U\Sigma V^\dagger.
\]
The Berry flux through a plaquette with oriented vertices
$0\rightarrow1\rightarrow2\rightarrow3\rightarrow0$ and the total Chern
number are
\begin{equation}
  F(\boldsymbol k)
  =-\operatorname{Im}\ln\det\!\left[
    \widetilde M_{01}\widetilde M_{12}
    \widetilde M_{23}\widetilde M_{30}
  \right],
  \qquad
  C=\frac{1}{2\pi}\sum_{\boldsymbol k}F(\boldsymbol k).
  \label{eq:FHS-Chern}
\end{equation}

For the Wilson loop, we multiply the unitary link matrices along one cBZ
direction, chosen here to be parallel to $\boldsymbol b_1$, at fixed
transverse momentum $k_\perp$:
\[
  W(k_\perp)
  =\prod_i\widetilde M_{i,i+1}(k_\perp).
\]
The eigenphases of $W(k_\perp)$ are the Wannier charge centers
$\theta_n(k_\perp)$. We track their continuous evolution as a function of
$k_\perp$.

When $U(1)_v$ is preserved, $h_{\mathrm{HF}}$ and the occupied projector are
block diagonal in the two valleys. We then construct the occupied subspace
and the overlap matrices independently in the $K$ and $K'$ blocks, avoiding
an arbitrary mixing of exactly or nearly degenerate valley states. This
yields the valley-resolved Berry fluxes $F^{(\tau)}(\boldsymbol k)$, Chern
numbers $C_\tau$, and Wilson-loop phases
$\theta_n^{(\tau)}(k_\perp)$. The total and spin/valley Chern numbers are
$C=C_K+C_{K'}$ and $C_s=(C_K-C_{K'})/2$, respectively. The integer $C_s$ is
well defined only when $U(1)_v$ is preserved. For a
time-reversal-invariant state, $C_K=-C_{K'}$, and the parity of $C_K$ agrees
with the $\mathbb Z_2$ invariant obtained from the Wilson-loop flow. When
$U(1)_v$ is broken by intervalley coherence, only the topology of the full
occupied subspace, including its total Chern number and Wilson-loop
spectrum, is gauge invariant.
\begin{figure}
    \centering
    \includegraphics[width=0.75\linewidth]{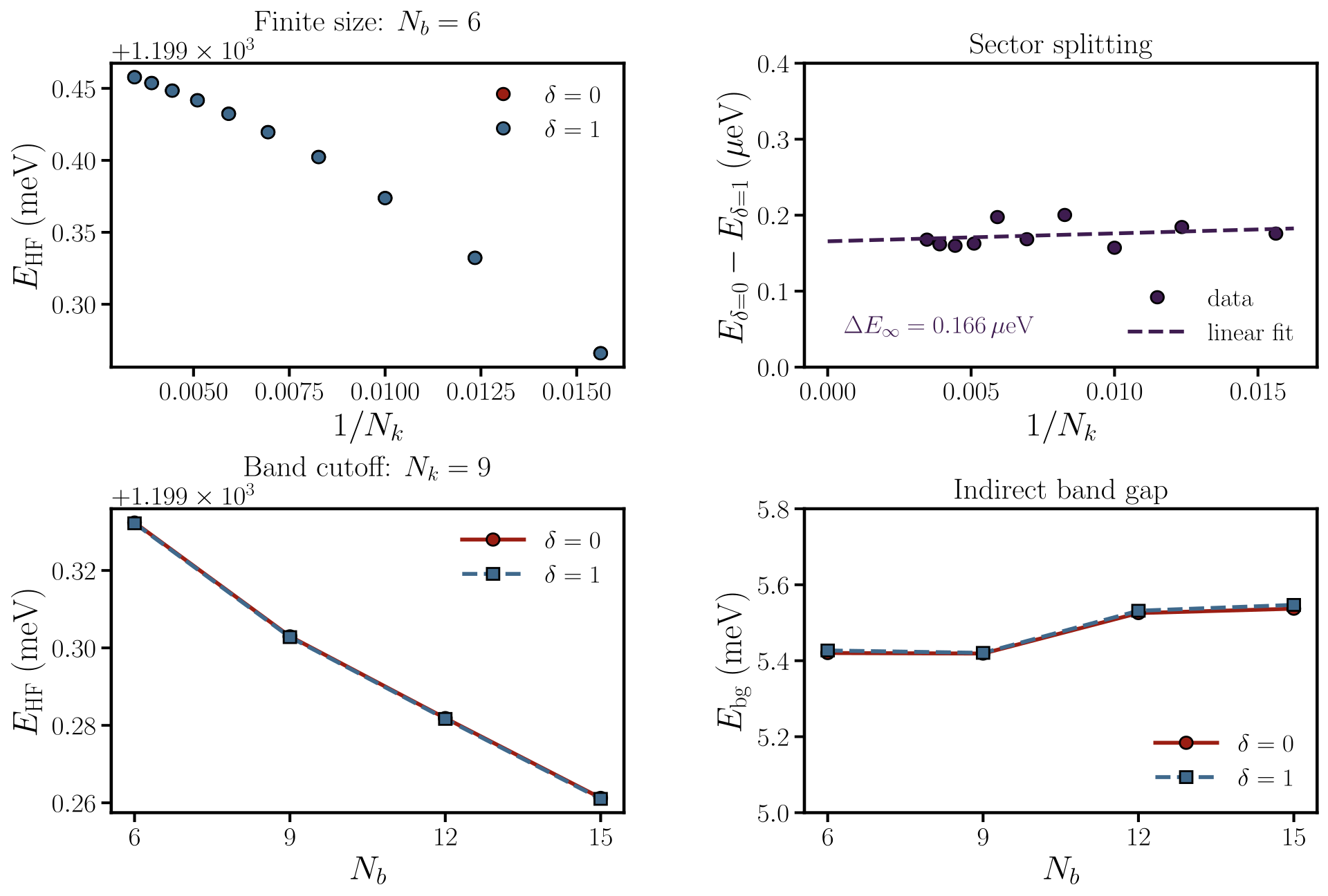}
    \caption{Finite-size and band-cutoff convergence of the $\delta=0$ and $\delta=1$ QSHC states. The calculations are performed at $\theta=5.25^\circ$,
  $\epsilon=20$, and $V_z=10\,\mathrm{meV}$. The upper panels show the Hartree–Fock energies and sector splitting $E_{\delta=0}-E_{\delta=1}$ versus $1/
  N_k$ for $N_b=6$; linear extrapolation gives finite positive splitting $\Delta E_\infty=0.17\mu\mathrm{eV}$. The lower panels show the Hartree–Fock energies and indirect band
  gaps versus the number of retained bands $N_b$ at $N_k=9$. The $\delta=1$ state remains energetically favored, while the indirect gap is stable against
  increasing the band cutoff.}
    \label{fig:SM_energies}
\end{figure}

To assess the numerical robustness of the relative stability between the
  $\delta=0$ and $\delta=1$ QSHC states, we examine both the finite-size and
  band-cutoff dependence of their Hartree--Fock energies. Here we define $N_{\boldsymbol{k}}^2$ as the number of points in the Monkhorst--Pack mesh and $N_b$ as the number of projected crystal bands used in SCHF calculations. At fixed $N_b=6$,
  the absolute Hartree--Fock energy varies appreciably with $N_k$, whereas the
  energy splitting,
  \begin{equation}
      \Delta E = E_{\delta=0}-E_{\delta=1},
  \end{equation}
  remains positive and is confined to approximately
  $0.16$--$0.20\,\mu\mathrm{eV}$. A linear extrapolation in $1/N_k$ gives
  $\Delta E_{\infty}\simeq 0.166\,\mu\mathrm{eV}$, indicating that the
  $\delta=1$ state remains energetically favored in the thermodynamic limit.
  At fixed $N_k=9$, increasing the number of retained bands from $N_b=6$ to
  $N_b=15$ shifts the absolute Hartree--Fock energy but does not reverse the
  ordering of the two states. Meanwhile, the indirect band gap remains stable
  at approximately $5.4$--$5.5\,\mathrm{meV}$. These results show that the
  preference for the $\delta=1$ registry is robust against both momentum-space
  finite-size effects and the band cutoff, although the associated energy scale
  is only of order $10^{-1}\,\mu\mathrm{eV}$ and therefore requires tightly
  converged calculations. In the main text, we use $N_{\boldsymbol{k}} = 9$-$15$ and $N_{b} = 6$ and set the energy convergence threshold to be $10^{-5}\text{meV}$, hence the energy splitting here remains small but physically resolved.
\begin{figure}
    \centering
    \includegraphics[width=0.98\linewidth]{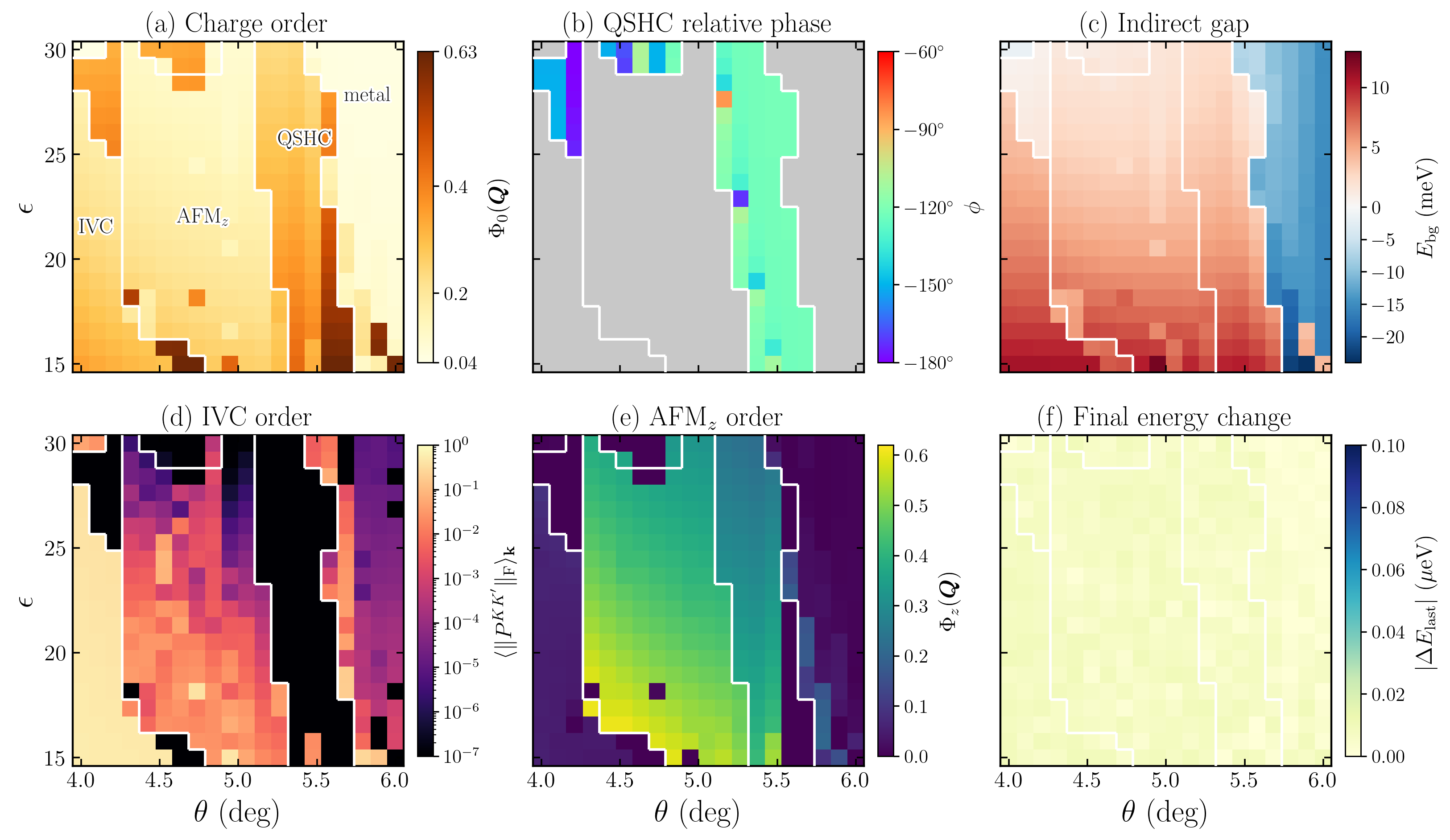}
    \caption{  Hartree--Fock diagnostics across the $\theta$--$\epsilon$ phase diagram at
  $V_z=10\,\mathrm{meV}$.
  (a) Intravalley charge-order order parameter $\Phi_0(\boldsymbol{Q})$.
  (b) Relative phase $\phi$ between the valley-resolved $\sqrt{3}\times\sqrt{3}$
  density component within the QSHC region; gray denotes points outside this region.
  (c) Indirect band gap $E_{\mathrm{bg}}$.
  (d) Intervalley-coherence (IVC) order defined as the Frobenius norm of the intervalley blocks, see main text for definition.
  (e) Out-of-plane antiferromagnetic order $\Phi_z(\boldsymbol{Q})$.
  (f) Absolute change in the Hartree--Fock energy during the final relaxation step.
  White lines indicate the boundaries between the metal, IVC crystal,
  AFM$_z$, and QSHC phases.}
    \label{fig:sm_hfpd}
\end{figure}

We supplement the analysis of the $\epsilon$--$\theta$ Hartree--Fock phase
  diagram presented in the main text by showing several complementary order
  parameters used to identify the phases. As discussed in the main text, all
  insulating phases exhibit a robust intravalley charge modulation
  $\Phi_0(\boldsymbol{Q})$, confirming that
  $\sqrt{3}\times\sqrt{3}$ translational-symmetry breaking persists throughout
  the crystal regime in Fig.~\ref{fig:sm_hfpd}(a)(c). We find that the intravalley charge order is stronger than the other symmetry-breaking phase, since it becomes the only channel to support the gap. In Fig.~\ref{fig:sm_hfpd}(b), we also evaluate the relative phase $\phi$ between
  the valley-resolved crystal harmonics within the QSHC region and find
  $\phi\simeq-120^\circ$ (equivalently $120^\circ$, depending on the phase
  convention), consistent with the theoretical analysis in the main text. A few
  points near the phase boundaries exhibit larger deviations from the locked
  $120^\circ$ value. Since these solutions satisfy the prescribed convergence
  criterion, we attribute these deviations to relaxation of the relative crystal
  phase induced by enhanced intervalley interactions and competition between
  nearly degenerate states. In the IVC region, the intervalley off-diagonal block of
  the density matrix is strongly enhanced in Fig.~\ref{fig:sm_hfpd}(d), whereas the AFM$_z$ region is
  characterized by a large out-of-plane staggered component
  $\Phi_z(\boldsymbol{Q})\sim10^{-1}$ in Fig.~\ref{fig:sm_hfpd}(e). A small residual IVC component, ranging
  from approximately $10^{-5}$ to $10^{-2}$, remains in parts of the AFM$_z$
  region. This residual coherence is sufficiently weak that the numerically
  extracted valley-resolved Chern number remains quantized. In the $\delta=1$ QSHC region, the IVC component is further
  suppressed, while the AFM$_z$ component is substantially weaker. The final
  energy changes across the phase diagram remain below the chosen convergence
  threshold, as shown in Fig.~\ref{fig:sm_hfpd}(f). We have additionally examined
  the valley-polarized branch and find that it lies approximately
  $5\,\mathrm{meV}$ above the states shown in the phase diagram. It
  therefore does not constitute a competitive ground-state candidate in this
  parameter regime.
\end{document}